\documentclass[11pt]{amsart}

\usepackage{graphicx}
\usepackage{dcolumn}
\usepackage{bm}
\usepackage{mathrsfs}
\usepackage{amsrefs}
\usepackage{hyperref}
\usepackage{amssymb}
\usepackage{amsaddr}

\newcommand{\mcal}[1]{\mathcal{#1}}

\newcommand{\inn}[1]{\left<#1\right>}
\newcommand{\mbb}[1]{\mathbb{#1}}
\newcommand* {\ee}{\ensuremath{\mathrm{e}}}

\newcommand{\N}{\mbb{N}}
\newcommand{\C}{\mbb{C}}

\DeclareMathOperator{\id}{id}

\newtheorem{thm}{Theorem}[section]

\begin{document}


\title{On the limit relation for the quantum relative entropy}
\author{J. Z. Bern\'ad}
\address{Institut f\"{u}r Angewandte Physik, Technische Universit\"{a}t Darmstadt, D-64289 Darmstadt, Germany}
\email{zsolt.bernad@physik.tu-darmstadt.de}
\author{A. B. Frigyik}
\address{Institute of Mathematics and Informatics, University of P\'ecs, H-7624 P\'ecs, Hungary}
\email{afrigyik@gamma.ttk.pte.hu}
\date{\today}


\date{\today}

\begin{abstract}
Recently, Di\'osi {\it et al.} \cite{Diosi} 
introduced a simple, yet very interesting
model for reservoirs, in order to study the relationship between
thermodynamic entropy production of a system and the corresponding von Neumann entropy (or informatic
entropy, as it was called by the authors). They came up with a conjecture about
the asymptotic behaviour of the state introduced in the model. The
conjecture was proven later by Csisz\'ar {\it et al.} \cite{Petz}. 
In this
paper we give a different interpretation of the original question and extend the result to a collection of more general states.
\end{abstract}

\maketitle


\section{Introduction}

In a thermodynamic model, Di\'osi {\it et al.} \cite{Diosi} considered an elementary model of a reservoir,
in which the thermodynamic entropy production due to the application of an
external field equals the change of 
the informatic entropy. Let us recapitulate the main ideas of this model. It has been assumed that this reservoir is composed by
$n$ distinguishable particles with finite energy levels. The time evolution of these particles is governed by the 
non-interacting Hamilton operator
\begin{equation}
 H_R=H\otimes I^{\otimes (n-1)}+I \otimes H \otimes I^{\otimes (n-2)}+ \dots + 
 I^{\otimes (n-1)} \otimes H. \nonumber
\end{equation}
Each particle is considered to be in a Gibbs state with a fixed inverse temperature $\beta$ and the state of the reservoir
is assumed to be completely uncorrelated:
\begin{equation}
 \rho_R=\rho^{\otimes n}, \quad \rho=\frac{\ee^{-\beta H}}{\mathrm{Tr} \{ \ee^{-\beta H} \}}.
 \label{Gibbs}
\end{equation}
An external field is applied instantaneously on the reservoir, i.e a unitary
evolution takes place on a much smaller time scale than the time
scale of the reservoir's evolution, and it is assumed that only one of the particles changes its state to $\sigma$:
\begin{equation}
 \sigma=U \rho U^*,\quad [U, H] \neq 0, \nonumber
\end{equation}
where $U$ is a unitary operator. This results in $n$ possible reservoir states $\rho'_R$
\begin{equation}
 \sigma \otimes \rho^{\otimes (n-1)},\, \rho \otimes \sigma \otimes \rho^{\otimes (n-2)},\, \dots,\, \rho^{\otimes (n-1)} \otimes \sigma, \nonumber
\end{equation}
depending which particle changed its state. The reservoir changes its mean energy by
\begin{equation}
 \Delta E = \mathrm{Tr}\{H_R \rho'_R\}-\mathrm{Tr}\{H_R \rho_R\}=\beta^{-1} S(\sigma || \rho)>0 \nonumber
\end{equation}
where $S(\sigma || \rho)$ is the relative entropy and we used the relation 
$\log \rho=-\beta H- \log (\mathrm{Tr} \{ \ee^{-\beta H} \})$(see
Eq. \eqref{Gibbs}) together with the 
fact that the von Neumann entropy is unitarily invariant:
\begin{equation}
 S(\sigma)=-\mathrm{Tr}\{\sigma \log (\sigma)\}=-\mathrm{Tr}\{U\rho U^* \log (U\rho U^*)\}=S(\rho). \nonumber
\end{equation}
If this mean energy 
is dissipated in the reservoir, then the thermodynamic entropy production is
equal to $S(\sigma || \rho)$. But 
the informatic entropy production is zero:
\begin{equation}
  S(\rho'_R)- S(\rho_R)=0. \nonumber
\end{equation}
The authors of Ref.~\cite{Diosi} 
needed a new state $R_n$ instead of $\rho'_R$ such that
\begin{equation}
 \mathrm{Tr}\{H_R \rho'_R\}=\mathrm{Tr}\{H_R R_n\} \quad \mathrm{and} \quad S(R_n)-S(\rho'_R) \rightarrow S(\sigma||\rho) \nonumber
\end{equation}
in the thermodynamic limit $n \to \infty$. It was conjectured that for 
\begin{equation}
R_n=\frac{\sigma \otimes \rho^{\otimes (n-1)}+\rho \otimes \sigma \otimes \rho^{\otimes (n-2)}+\dots+ \rho^{\otimes (n-1)} \otimes \sigma}{n} 
\nonumber 
\end{equation}
the informatic entropy production rate is equal to the relative entropy $S(\sigma||\rho)$ in the thermodynamic limit. This conjecture was proved
by Csisz\'ar {\it et al.} \cite{Petz}. The proof generalizes the conjecture of Ref.~\cite{Diosi} for arbitrary density matrices $\sigma$ and $\rho$,
which are not related any more by a unitary transformation, i.e. possibly
$S(\rho'_R)\neq S(\rho_R)$, and shows that
\begin{eqnarray}
 &&\lim_{n \to \infty} \Big(S(R_n)-S(\rho'_R)\Big)= S(\sigma||\rho), \label{entconv} \\
 &&\forall \rho'_R \in \{\sigma \otimes \rho^{\otimes (n-1)},\, \rho \otimes \sigma \otimes \rho^{\otimes (n-2)},\, \dots,\, 
 \rho^{\otimes (n-1)} \otimes \sigma\}. \nonumber
\end{eqnarray}

This whole problem can be viewed from a different angle. Let us consider the approach of M. J. Donald \cite{Donald} and calculate the Helmholtz
free energy $F$ of the reservoir in the state $R_n$ and $\rho_R=\rho^{\otimes n}$. The difference between them is given by
\begin{equation}
 F(R_n)-F(\rho^{\otimes n})=\beta^{-1} S(R_n||\rho^{\otimes n}). \nonumber
\end{equation}
One can simply compute
\begin{eqnarray}
 S(R_n||\rho^{\otimes n})=-S(R_n)+S(\rho'_R)+S(\sigma||\rho). \nonumber
\end{eqnarray}
Thus Eq. \eqref{entconv} is equivalent to
\begin{equation}
 S(R_n||\rho^{\otimes n}) \rightarrow 0 \quad \mathrm{as} \quad n \to \infty, \nonumber
\end{equation}
which yields
\begin{equation}
 F(R_n)-F(\rho^{\otimes n}) \rightarrow 0 \quad \mathrm{as} \quad n \to \infty. \nonumber
\end{equation}

This means that in the thermodynamic limit the reservoir changed its equilibrium state to another equilibrium state. A natural question is to find 
all these thermodynamically equivalent states. In 
case 
when only one particle changed its Gibbs state $\rho$ to an arbitrary 
state $\sigma$, the proof in Ref.~\cite{Petz} shows that the state $R_n$ 
-- where the randomization is uniform, i.e. each individual state has the same
probability to change to state $\sigma$ -- is  
an equilibrium state equivalent to the inital state $\rho^{\otimes n}$. 

In this paper we consider a general case: 
We suppose that the randomization is not uniform, hence 
we consider the state 
\[
 \rho_n = a_{n,1} \sigma \otimes \rho^{\otimes (n-1)}+ a_{n,2} \rho \otimes
 \sigma \otimes \rho^{\otimes (n-2)} +\dots 
 + a_{n,n} \rho^{\otimes (n-1)} 
 \otimes \sigma, \label{rhon}
\]
where the matrix $(a_{i,j})$ contains the changing weights for all $n$, such that $a_{n,i} \in [0,1]$ when $i \leqslant n$ otherwise $a_{n,i}=0$ 
and $\sum^n_{i=1}a_{n,i}=1$. This construction allows us to study the
thermodynamic limit $n \to \infty$. The main question is about the properties of this matrix 
that assure the following limit:
\begin{equation}
S(\rho_n||\rho^{\otimes n}) \rightarrow 0 \quad \mathrm{as}  \quad n \to \infty. \nonumber 
\end{equation}

In general, a matrix $(a_{i,j})$ is regular or T-matrix (see Ref.~\cite{Cooke}), if the following conditions are satisfied:
\begin{eqnarray}
 &&\sup_{n \in \mathbb{N}^+} \sum^{n}_{i=1} |a_{n,i}|< \infty, \quad \lim_{n \to \infty} \sum^{n}_{i=1} a_{n,i}=1, \nonumber \\
 &&\lim_{n \to \infty} a_{n,j}=0,\quad \forall j\in \mathbb{N}^+. \nonumber
\end{eqnarray}
It is strongly regular if it is regular and (see Ref.~\cite{Jardas})
\begin{equation}
 \lim_{n \to \infty} \sum^\infty_{j=1} |a_{n,j+1}-a_{n,j}|=0. \nonumber 
\end{equation}

We are going to give an analytical proof of the  following statement:
\begin{thm}\label{thm1}
 Let $\rho$ be a Gibbs state and $\sigma$ a state such that $\mathrm{supp}(\sigma) \subseteq \mathrm{supp}(\rho)$. Then for every
 strongly regular matrix $(a_{i,j})$, where $a_{i,j}\ge 0$ for all
 $i,j\in\mathbb{N}^+$ and $\sum_{i=1}^n a_{n,i}=1$ for all $n\in \mathbb{N}^+$,
 and 
\[
 \rho_n = a_{n,1} \sigma \otimes \rho^{\otimes (n-1)}+ a_{n,2} \rho \otimes
 \sigma \otimes \rho^{\otimes (n-2)} + \dots 
 + a_{n,n} \rho^{\otimes (n-1)} 
 \otimes \sigma 
\]
we have
\begin{equation}
 \lim_{n \to \infty} S(\rho_n||\rho^{\otimes n})=0. \nonumber
\end{equation}
\end{thm}


The entries of the strongly regular matrix $(a_{i,j})$ must be restricted to the
interval $[0,1]$ in order for the state in \eqref{rhon} to be a 
real physical state of the reservoir. Thus, Theorem \ref{thm1} is capable to identify a large class of states for which the Helmholtz 
free energy in the thermodynamic limit is the same as for the equilibrium state $\rho^{\otimes n}$, where $n \in \mathbb{N}^+$. The presented 
proof combines the idea that comes from Ref.~\cite{Petz} with results from ergodic theory \cite{Krengel}.

\section{GNS approach}

Let us start with Eq. \eqref{rhon}: Applying the same idea that was used in
Ref.~\cite{Petz} we can rewrite $\rho_n$ in a more suitable form. Let $X=\rho^{-1/2}\sigma\rho^{-1/2}$, where $\rho$ is invertible 
(see definition in \eqref{Gibbs}). Then we have
\[
\rho_n = \left(\rho^{\frac{1}{2}}\right)^{\otimes n} \Big( a_{n,1} X \otimes
I^{\otimes (n-1)} + a_{n,2} I \otimes X \otimes I^{\otimes (n-2)} 
+\cdots + a_{n,n} I^{\otimes (n-1)} \otimes X\Big)
\left(\rho^{\frac{1}{2}}\right)^{\otimes n}. 
\]

Here, we cannot use directly the mean ergodic theorem that was applied in
Ref.~\cite{Petz} or Ref.~\cite{Ohya} (see Proposition 1.17 there). 
Nevertheless, we will follow a very similar path to that which was given in the
proof of the aforementioned proposition. The algebra underlying our problem is the same as the proposition's
and hence we are going to adopt the notation used in Ref.~\cite{Ohya}. As we have
already stated, the reservoir consists of
particles with finite energy levels, and we denote by $d$ the dimension of the
Hilbert space underlying their quantum mechanical 
description. Let $M_d$ denote the set of all $d \times d$ complex
matrices acting on this Hilbert space. Since $M_d$ is a $C^*$-algebra
with identity, we consider the family of $C^*$-algebras 
$\{M_{d^n}=(M_d)^{\otimes n}| n \in \mathbb{N}^+\}$ the $n$-fold algebraic
tensor products of $M_d$. If we identify $M_{d^n}$ with $M_{d^n}\otimes I$ in 
$M_{d^{n+1}}$, where $I=I_d$ denotes the identity operator on the $d$ dimensional
Hilbert space, we can consider
$M_{\infty}=\bigcup_{n=1}^\infty M_{d^n}$, the inductive limit $\ast$-algebra of $\{M_{d^n}\}$, and its completion
$A_{\infty}$. In other words,
\[
A_{\infty}=\overline{\bigotimes_{i=1}^\infty A_i},
\]
where $A_i=M_d$ for all $i\in\mathbb{N}^+$ 
(for the details see Ref.~\cite{Sakai}, Section 1.23). The family of faithful states 
(see Eq. \eqref{Gibbs}) $\{\varphi_n(x)=\mathrm{Tr} \{\rho^ {\otimes n} x\}| x
\in M_{d^n}, \, n \in \mathbb{N}^+\}$, are functionals on
the particular $C^*$-algebras, which extend to a faithful state
$\varphi_{\infty}$ on $A_{\infty}$. The GNS representation
of $A_{\infty}$ with respect to the state $\varphi_{\infty}$ results in a triplet $(\pi, \mathcal{H}, \Omega)$, where $\mathcal{H}$ is the constructed
Hilbert-space, $\pi$ is a $*$-homomorphism of $A_{\infty}$ into the set of all bounded operators $\mathcal{B}(\mathcal{H})$ on $\mathcal{H}$, 
and $\Omega$ is the cyclic vector for $\pi$. We generate the von Neumann algebra $\mathcal{M}=\pi(A_{\infty})''$ by taking the bicommutant of $\pi(A_{\infty})$
and thus we make sure that all weakly or strongly convergent series have their limit point in $\mathcal{M}$.

Just like in Ref.~\cite{Ohya} we are going to denote by $\gamma$ the
right shift on $A_{\infty}$ which is defined for $x_1 \otimes x_2 \otimes \dots \otimes x_n \in M_{d^n}$ as
\begin{equation}
 \gamma(x_1 \otimes x_2 \otimes \dots \otimes x_n)=I\otimes x_1 \otimes x_2 \otimes \dots \otimes x_n \in M_{d^{n+1}}. \nonumber
\end{equation}
As a result, $\rho_n$ can be
written as 
\begin{equation}
\rho_n=\left(\rho^{1/2}\right)^{\otimes n}\left( \sum_{i=0}^{n-1} a_{n,i+1}\gamma^i(X)
\right) \left(\rho^{1/2}\right)^{\otimes n} \nonumber
\end{equation}
It is obvious that $\varphi_{\infty}$ is invariant under $\gamma$. The extension $\varphi$ of the state $\varphi_{\infty}$
to $\mathcal{M}$ is 
\begin{equation}
 \varphi(a)=\langle \Omega, a \Omega \rangle, \,\, a \in \mathcal{M}, \nonumber
\end{equation}
and the right shift $\gamma$ can be also extended such that $\varphi \circ \gamma = \varphi$. Now, the only
missing link in this approach is the connection between the relative entropy $S(\rho_n||\rho^{\otimes n})$ and the state $\varphi$.
Let us recall the Belavkin-Staszewski entropy and its relation to the relative entropy (see Ref.~\cite{Petz} and Proposition 7.11 in Ref.~\cite{Ohya})
\[
  S\left(\rho_n||\rho^{\otimes n}\right) \leqslant
  S_{BS}\left(\rho_n||\rho^{\otimes n}\right) 
  =\mathrm{Tr} \Big\{\rho^{\otimes n} \eta\left((\rho^{-1/2})^{\otimes n} \rho_n (\rho^{-1/2})^{\otimes n} \right)  \Big \} 
\]
where $\eta(t)=-t\log t$ and $\mathrm{supp}(\sigma) \subseteq \mathrm{supp}(\rho)$. One can simply realize that the right hand side 
of the above equation is nothing else than the state $\varphi_n$ and therefore with the help of the GNS construction  
we get
\begin{equation}
 S\left(\rho_n||\rho^{\otimes n}\right) \le \inn{\Omega,\eta\left( \sum_{i=0}^{n-1} a_{n,i+1}\gamma^i(\mathcal{X}) \right)\Omega}, \label{todo}
\end{equation}
where $\Omega$ is the cyclic vector and $\mathcal{X} \in \mathcal{M}$ is the embedding of $X=\rho^{-1/2}\sigma\rho^{-1/2}$ into $\mathcal{M}$.

\section{Proof of Theorem \ref{thm1}}

We would like to prove first that $\sum_{i=0}^{n-1}
a_{n,i+1}\gamma^i(\mathcal{X})\to \id_{\mcal{H}}$ strongly. In order to do that we need the
following theorem 
(Theorem \ref{thm4}) from
Ref.~\cite{Jardas} (see Ref.~\cite{Yosida} pg.~213 for the
original theorem). For convenience of the reader we repeat a slightly modified
version of the statement here.


\begin{thm}\label{thm4}
  Let $Y$ be a Banach space and let $T:Y\to Y$ be a continuous linear operator
  such that the family of operators $\{T^n,n\in \mathbb{N}\}$ is power bounded. Further, let $A=(a_{n,i})(n,i\in \mathbb{N}^+)$ be a strongly regular matrix and
\[
 T_n=\sum_{i=1}^\infty a_{n,i}T^i,\quad (n\in \mathbb{N}^+).
\]
Then the strong limit $T_0=\lim_{n\to \infty} T_n$ exists and it is a projection on $Y$ onto the subspace $\mcal{F}=\{y\in Y:Ty=y\}$ of all
  fixed points of $T$. Moreover, projection $T_0$ does not depend on the choice
  of the strongly regular matrix $A$.
\end{thm}

The GNS construction provides us with a Hilbert space $\mcal{H}$ which is a
reflexive Banach space 
and for the rest of the paper we are going to adopt the proof of Proposition
1.17 in Ref.~\cite{Ohya}. Let $V:\mcal{H}\to \mcal{H}$ be defined by
\begin{equation}
Va\Omega= \gamma(a)\Omega, \quad \forall a\in\mcal{M}. \nonumber 
\end{equation}
The map $V:\mcal{H}\to \mcal{H}$ is an isometry on $\mcal{A}_\infty \Omega$: 
\begin{multline*}
\inn{Va\Omega,Vb\Omega}= \inn{\gamma(a)\Omega,\gamma(b)\Omega} = \inn{(I\otimes
  a)\Omega, (I\otimes b)\Omega}\\ = \inn{\Omega, (I\otimes a^*)(I\otimes b)
  \Omega} = \inn{\Omega, (I\otimes a^*b)\Omega}= \varphi_\infty(I\otimes
  a^*b)\\ = \varphi_{\infty}(I)\varphi_\infty(a^*b) = \varphi_\infty(a^*b)=
  \inn{\Omega, a^*b\Omega} = \inn{a\Omega, b\Omega}.
\end{multline*}
Here, we used the fact that $\varphi_{\infty}\restriction_{M_{d^n}}=\varphi_n$.
Since $V$ is an isometry and $\overline{\pi(\mcal{A}_\infty)\Omega}=\mathcal{H}$,
$\{V^n,n\in\N\}$ is a power bounded family on
$\mcal{H}$, which means that we can apply Theorem \ref{thm4} 
to our situation, where $A=(a_{i,j})$ is a strongly regular matrix. The isometry $V$ plays the role of $T$, hence
we need to find the fixed point subspace of $V$. Since the matrix $A$ does not
play any role in finding this subspace, the proof of Proposition 1.17 in
Ref.~\cite{Ohya} works verbatim and therefore the dimension of $\mcal{F}$ is one. Since 
\begin{equation}
V\id_{\mcal{A}_\infty}\Omega=\gamma\left(\id_{\mcal{A}_\infty}\right)\Omega=
\gamma(I^{\otimes\infty})\Omega= I^{\otimes\infty}\Omega= \id_{\mcal{A}_\infty}\Omega, \nonumber
\end{equation}
the $\Omega$ is in $\mcal{F}$ and hence $\mcal{F}=\C\Omega$. 

If $b'\in\mcal{M}'$, where $\mathcal{M}'$ is the commutant of $\mathcal{M}$, and $S_n(a)=\sum_{i=0}^{n-1} a_{n,i+1}\gamma^i(a)$,
 $a \in \mathcal{M}$, then 
\[
S_n(a)b'\Omega=b'S_n(a)\Omega=b'\sum_{i=0}^{n-1} a_{n,i+1}\gamma^i(a)\Omega 
=b'\sum_{i=0}^{n-1} a_{n,i+1}V^ia\Omega \to b'Ea\Omega, 
\]
by Theorem \ref{thm4}, where $E$ is the projection of $\mcal{H}$ 
 onto $\mcal{F}$: $E(\xi)=\inn{\Omega,\xi}\Omega$ for all $\xi\in \mcal{H}$.
Thus
\[
b'Ea\Omega = b'\inn{\Omega,a\Omega}\Omega= \varphi(a)b'\Omega.
\]
We know from the proof of Proposition 1.17 in Ref.~\cite{Ohya} that
$\Omega$ is cyclic for $\mcal{M}'$, as well, hence $S_n(a)\to
\varphi(a)\id_{\mcal{H}}$ strongly. If we choose $a=\mathcal{X}$ then 
\[
\varphi(\mathcal{X})=\mathrm{Tr} \{\rho \rho^{-1/2} \sigma \rho^{-1/2}\}=1.   
\]
Hence $S_n(\mathcal{X})\to \id_{\mcal{H}}$ strongly. 

We can finish the proof now. 
Operator $\mathcal{X}$ is positive and the function $\eta(t)=-t\log t$ is
continuous on $[0,\infty)$. Since the continuous functional calculus preserves
the strong convergence, we obtain that $ \lim_{n \to \infty} \eta
\left(S_n(\mathcal{X}) \right)=0$ strongly. Therefore, with the aid of Eq.~\eqref{todo}
\[ 
\lim_{n \to \infty} \inn{\Omega,\eta\left( \sum_{i=0}^{n-1} a_{n,i+1}\gamma^i(\mathcal{X}) \right)\Omega}=0 \quad 
 \Rightarrow \quad \lim_{n \to \infty} S\left(\rho_n||\rho^{\otimes n}\right)=0
\]
and the proof is complete. 

Similarly, one can also obtain for 
\[
 \rho_{n,k} = a_{n,1} \sigma^{\otimes k} \otimes \rho^{\otimes (n-k)}+ a_{n,2} \rho \otimes \sigma^{\otimes k} \otimes \rho^{\otimes (n-k-1)} 
 +\dots+ a_{n,n-k+1} \rho^{\otimes (n-k)} \otimes \sigma^{\otimes k} 
\]
with $\mathcal{X}=\rho^{-1/2}\sigma^{\otimes k}\rho^{-1/2}$, $\sum^{n-k+1}_{i=1} a_{n,i}=1$ and fixed $k$ the limit relation
\begin{equation}
 \lim_{n \to \infty} S(\rho_{n,k}||\rho^{\otimes n})=0. \nonumber
\end{equation}
This result extends a similar statement made in Ref.~\cite{Petz} and
further extends the set of states identified in Theorem \ref{thm1}.

\section*{Acknowledgement}

This work was supported by the Deutscher Akademischer Austauschdienst (M\"OB-DAAD project no. 65049).

\nocite{*}

\end{document}